\documentclass[11pt]{article}

\usepackage{aas_macros,amsmath,amssymb,comment,cite,esint,graphicx,mathtools}
\usepackage[margin=.8in,letterpaper]{geometry}
\usepackage[colorlinks=true]{hyperref}
\usepackage[affil-it]{authblk}
\usepackage{subcaption}
\usepackage[utf8]{inputenc}
\usepackage{mathrsfs}
\usepackage{appendix}
\usepackage{amssymb}
\usepackage{float}
\usepackage{color}
\usepackage{cite}
\usepackage{hyperref}
\hypersetup{pageanchor=false}
\usepackage{indentfirst}
\usepackage{url}
\usepackage{float}
\usepackage{caption}
\usepackage[numbers,square,comma,sort&compress,merge]{natbib}
\usepackage{esint}
\usepackage{overpic}
\usepackage{graphicx}
\usepackage{epsf,amsmath,bbold,amsfonts,stmaryrd}
\usepackage{textcomp}
\usepackage{ulem}
\usepackage{cancel}
\usepackage{tikz}

\numberwithin{equation}{section}
\let\originalleft\left
\let\originalright\right
\renewcommand{\left}{\mathopen{}\mathclose\bgroup\originalleft}
\renewcommand{\right}{\aftergroup\egroup\originalright}
\mathcode`\*="8000
{\catcode`\*=\active\gdef*{\mathclose{}\,\mathopen{}}}

\newcommand{\td}[1]{\tilde{#1}}

\newcommand{\be}{\begin{equation}}
\newcommand{\ee}{\end{equation}}
\newcommand{\bea}{\setlength\arraycolsep{2pt} \begin{eqnarray}}
\newcommand{\eea}{\end{eqnarray}}
\newcommand{\nn}{\nonumber}

\newcommand{\mE}{{\mathcal E}}

\newcommand{\EM}{{\text{em}}}
\newcommand{\ion}{{\text{ion}}}
\newcommand{\te}{{\text{e}}}

\def\a{\alpha}
\def\b{\beta}

\def\D{\Delta}
\def\f{\frac}
\def\g{\gamma}

\def\lm{\lambda}

\def\m{\mu} 
\def\n{\nu} 
\def\nn{\nonumber}

\def\p{\phi} 
\def\td{\tilde} 

\def\s{\sigma}
\def\S{\Sigma}
\def\t{\theta}
\def\T{\Theta}

\def\be{\begin{equation}}
\def\ee{\end{equation}}
\def\bag{\begin{aligned}}
\def\eag{\end{aligned}}
\def\bea{\begin{eqnarray}}
\def\eea{\end{eqnarray}}
\def\ba{\begin{array}}
\def\ea{\end{array}}

\def\bc{\begin{center}}
\def\ec{\end{center}}

\begin{document}
\title{Imaging thick accretion disks and jets surrounding black holes}
	
\author{Zhenyu Zhang$^{1}$, Yehui Hou$^{1}$, Minyong Guo$^{2,3\ast}$, Bin Chen$^{1, 4, 5}$}
\date{}
	
\maketitle
\vspace{-10mm}

\begin{center}
{\it
$^1$School of Physics, Peking University, No.5 Yiheyuan Rd, Beijing
100871, P.R. China\\\vspace{4mm}

$^2$ Department of Physics, Beijing Normal University,
Beijing 100875, P. R. China\\\vspace{4mm}

$^3$ Key Laboratory of Multiscale Spin Physics, Ministry of Education, Beijing 100875, P. R. China\\\vspace{4mm}

$^4$Center for High Energy Physics, Peking University,
No.5 Yiheyuan Rd, Beijing 100871, P. R. China\\\vspace{4mm}

$^5$ Collaborative Innovation Center of Quantum Matter,
No.5 Yiheyuan Rd, Beijing 100871, P. R. China\\\vspace{2mm}
}
\end{center}

\vspace{8mm}

\begin{abstract}
Based on the horizon-scale magnetofluid model developed in \cite{Hou:2023bep}, we investigate the millimeter-wave images of a geometrically thick accretion disk or a funnel wall, i.e., the magnetofluid that encloses the base of the jet region, around a Kerr black hole. By employing the numerical method to solve the null geodesic and radiative transfer equations, we obtain the optical appearances at various observational angles and frequencies, generated by the thermal synchrotron radiation within the magnetofluid. For the thick disk, we specifically examine the impact of emission anisotropy on images, concluding that anisotropic synchrotron radiation could play an important role in the observability of the photon ring. For the funnel wall, we find that both the outflow and inflow funnel walls exhibit annular structures on the imaging plane. The outflow funnel wall yields a brighter primary image than the photon ring, whereas the inflow one does not. Based on our investigation, the inflow funnel wall model can not be ruled out by current observations of M87*.
\end{abstract}

\vfill{\footnotesize $\ast$ Corresponding author: minyongguo@bnu.edu.cn}

\maketitle

\newpage
\baselineskip 18pt
\section{Introduction}\label{sec1}
	
In recent years, the horizon-scale photographs of supermassive black holes within the millimeter wave band have been viewed with great anticipation for exploring black hole physics \cite{EventHorizonTelescope:2019dse,EventHorizonTelescope:2022wkp,Lu:2023bbn}. It is believed that the observed bright rings are due to the so-called photon ring formed by multi-level images caused by the strong gravitational lensing effect \cite{teo2003spherical, Falcke:1999pj, Gralla:2019drh, Broderick:2022tfu, Hou:2022gge}. However, since the observations are sensitive to the astronomical environments around black holes, the extent to which the currently photographed bright rings are correlated with the photon rings is still a controversial topic \cite{Gralla:2020pra,Younsi:2021dxe,Ozel:2021ayr,Glampedakis:2021oie}. Therefore, it is essential to investigate more deeply the physical properties of the astronomical environments and their contributions to the black hole images.

It is widely recognized that a supermassive black hole accretes hot, magnetized plasma, thereby forming a luminous accretion disk. The light emitted by the thermal synchrotron electrons within this disk is believed to be the most probable source in the frequency band of black hole image observations. Moreover, for a high-spin black hole, the electromagnetic energy can potentially power a relativistic jet \cite{Blandford:1977ds}. The base of the jet is enveloped by what is known as the ``funnel wall'', capable of generating substantial thermal synchrotron radiation. Consequently, both the disk and the jet can serve as light sources, producing horizon-scale images of black holes.

In the field of theoretical research, many previous studies mainly focus on the gravitational lensing effect stemming from spacetime geometry, where the light sources are modeled to be of simple structures and often as geometrically thin disks \cite{luminet1979image, Peng:2020wun, Zeng:2021dlj, Chael:2021rjo, Guerrero:2021ues, Gan:2021xdl,Li:2021ypw, Gyulchev:2021dvt, Hou:2022eev,Wang:2022yvi,Zhang:2022klr,Meng:2023htc,Wang:2023vcv,Huang:2023ilm, Zhang:2023okw, DeMartino:2023ovj, Hu:2023pyd, Chakhchi:2022fls, Zhu:2023kei}. However, for a low-luminosity active galactic nuclei (LLAGN) like the M87*, the horizon-scale accretion flow is mostly geometrically thick \cite{Ho:1999ss,EventHorizonTelescope:2019dse}. In this case, it is essential to have precise knowledge of various factors such as the streamlines, particle number density, electron temperature, and the magnetic field structure. This crucial physical information can be obtained by solving the general relativistic magnetohydrodynamic (GRMHD) equations. Given the analytical challenges posed by the GRMHD equations, researchers typically resort to numerical techniques to ascertain the physical properties of the magnetofluid \cite{McKinney:2012vh, Moscibrodzka:2015pda,EventHorizonTelescope:2019pgp,Chatterjee:2020eqc,Ressler:2020voz,Shen:2023nij,Fraga-Encinas:2023nsa,Wong:2023uin}. However, there are shortcomings in numerical simulation, for example, the relationship between the initial and final states of simulation is somewhat ambiguous, and the numerical approach is computationally demanding. 

To balance computationally expensive numerical simulations and over-simplified emission models, semi-analytic models have been utilized in some literature. The basic idea behind these models is to circumvent the GRMHD simulation and give the distributions of number density, temperature, and magnetic field in other ways. For example, most of these works considered radiatively inefficient accretion flow (RIAF) models \cite{Broderick:2005at,Broderick:2008qf,Broderick:2010kx,Pu:2016qak,Pu:2018ute,Vincent:2020dij,Chen:2021lvo,Vincent:2022fwj,Jiang:2023img, Deliyski:2024wmt}, where the vertically-averaged electron density and temperature have a nearly power-law dependence on radius \cite{Yuan:2003dc}. Another model that was often utilized is the magnetized torus solution \cite{Komissarov:2006nz}. This model, characterized by its purely toroidal configuration, was frequently adopted as the initial condition for GRMHD simulations. Interestingly, some studies have also considered the images of the torus \cite{Vincent:2014erw,Vincent:2019nni}. After establishing the emission profiles for the accreted plasma, black hole images can be obtained by using general relativistic radiative transfer (GRRT) methods.

Recently, we introduced a new analytical model of magnetofluids surrounding a rotating black hole within the framework of GRMHD\cite{Hou:2023bep}. Assuming that the fluid acceleration at the event horizon scale is dominated by gravitation, we derived a comprehensive set of explicit expressions for thermodynamics and the associated magnetic field. We showed that our model can effectively describe the morphology of both thick disks and jets at the horizon scale and provides a direct pathway for the study of black hole imaging. In the present work, given that our model provides the essential physical parameters for the emission profile of the magnetofluid within the GRMHD framework, we would like to utilize the model to study the optical appearances of accretion disks and funnel wall near black holes, by using a GRRT method to generate the images numerically. Our investigation for the images of the disk will not be limited to analyzing the characteristics of intensity maps under various observational angles and frequencies, but will specifically probe the influence of emission anisotropy on the images. 

The remaining parts of the paper are organized as follows. In sec. \ref{sec2}, we concisely review the analytical model for thick accretion disks and jets, developed in \cite{Hou:2023bep}.  In sec. \ref{sec3} we introduce thermal synchrotron radiation and radiative transport,  which are fundamental in magnetohydrodynamic imaging. In Sec. \ref{sec4}, we present the imaging results for accretion disks and jets, accompanied by comprehensive discussions. Sec. \ref{sec5} provides a summary and discussion of our work. In addition, we will work in the unit with $GM  = c = 4\pi\epsilon_0= 1$, where $M$ is the mass of the black hole.

\section{Analytic model for thick accretion disks and jets}\label{sec2}

Recently, we have proposed a new analytical model for the magnetohydrodynamic fluid in \cite{Hou:2023bep}, trying to capture key features of thick disks and funnel walls. Specifically, we considered the fluid to be electrically neutral and postulate complete ionization of the plasma into electrons and protons. We focused on high-temperature systems with ultra-relativistic electrons in the adiabatic limit. To simplify the analysis, we introduced a linear relationship between the temperatures of electrons and ions, represented by a constant factor, $z =T_{\text{ion}}/T_{\text{e}}$, and considered both the non-relativistic and ultra-relativistic limits of ions. With the relativistic Euler equations for ideal, adiabatic magnetofluids, we expressed the temperature and pressure as power functions of the particle number density, with the exponents determined by $z$. Then, we demonstrated that in the scenarios where both the sound speed and Alf\'{v}en velocity are sub-relativistic, the streamlines follow geodesics in the leading-order  approximation, namely, the ballistic approximation. Furthermore, for a stationary, axisymmetric, ballistic fluid configuration in Kerr spacetime, we derived explicit expressions for the fluid thermodynamics under specific conditions regarding the streamlines. Especially, one of the conditions leads to the conical solution, which is of particular importance in the study of thick disks and funnel walls. For the detailed discussions on the model, please refer to \cite{Hou:2023bep}. 

 In this study, we would like to use this model to investigate the images of thick accretion disks and funnel walls encircling Kerr black holes. In a Kerr spacetime, the metric reads
\bea\label{metric}
\mathrm{d}s^2=-\left(1-\frac{2r}{\Sigma}\right)\mathrm{d}t^2+\frac{\Sigma}{\Delta}\mathrm{d}r^2+\Sigma\mathrm{d}\theta^2+\left(r^2+a^2+\frac{2ra^2}{\Sigma}\mathrm{sin}^2\theta\right)\mathrm{sin}^2\theta\mathrm{d}\phi^2-\frac{4ra}{\Sigma}\mathrm{sin}^2\theta\mathrm{d}t\mathrm{d}\phi\, , 
\eea
in Boyer-Lindquist (BL) coordinates. Note that we have set $GM = 1$ where $M$ is the mass of the black hole. In Eq. (\ref{metric}) we have defined 
\bea
\Delta=r^2-2r+a^2\,,\quad \Sigma=r^2+a^2\cos^2\theta\,,
\eea
where $a$ is the spin parameter of the black hole. From $\Delta=0$, the outer and inner radii of the horizons of a Kerr black hole can be obtained as
\bea
r_\pm=1\pm\sqrt{1-a^2}\,.
\eea
In the model proposed in \cite{Hou:2023bep}, the event horizon radius $r_h \equiv r_+$ serves as the inner boundary of the fluid. When considering the outer boundary of the fluid, we take it to be at infinity. This means that accretion inflows can extend from infinity to the event horizon, and the outflows can propagate indefinitely. Furthermore, the plasma streamlines within the fluid move along geodesic trajectories. Following the conical solution in \cite{Hou:2023bep}, let us further consider the plasma maintaining a constant $\theta$ in its geodesic motion. In Kerr spacetime, the four-velocity of such a timelike particle can be expressed as
\bea\label{u}
&&u^t = E\bigg[1+\f{2r(r^2+a^2)}{\D\S}\bigg]-\f{2arL}{\D\S} \, , \nn \\
&&u^r = \s_r \f{\sqrt{R}}{\Sigma} 
\, , \quad u^{\t} = 0\, , \nn \\
&&u^{\p} = E\f{2ar}{\D\S}+L\f{\csc^2\t(\D-a^2\sin^2{\t})}{\D\S} \, ,
\eea
where 
\bea\label{LL}
L = \pm_L a\sqrt{E^2-1}\sin^2{\t} \, , \, \, Q = -a^2(E^2-1)\cos^4{\t}\, ,
\eea
and
\bea\label{RR}
R(r,\t)=\left(E^2-1\right)r^4 + 2r^3 + 2a^2\left(E^2-1\right)\cos^2{\t}r^2+2\left[\big(aE-L\big)^2+Q\right]r- a^2Q\, ,
\eea
with ``$\pm_L$'' denoting the sign of $L$. Here $E$, $L$ are the conserved energy and angular momentum per unit mass and $Q$ is the Carter constant per unit square mass. In addition, $\s_r=\pm1$ denotes the direction of motion, either outward or inward. Note that one can demonstrate that $R\ge0$ is always true outside the event horizon through simple algebraic computations. Then, from Eq. (\ref{LL}) we can find that $E\ge1$ must hold if a particle following a conical geodesic can reach infinity. In such circumstances, by solving the equation of particle number conservation along streamlines, the number density of the plasma and electron temperature are expressed as
\bea\label{nT11}
n_\te(r,\t) = n_\te(r_h,\t) \sqrt{\f{R(r_h,\t)}{R(r,\t)}} \,,
\eea
and 
\bea\label{nT22}
T_\te(r,\t)= \left\{
\begin{aligned}
	&T_\te(r_h,\t) \bigg[\f{R(r_h,\t)}{R(r,\t)}\bigg]^{\f{(1+z)}{3(2+z)}}\,, \quad &z\ll z_c\\
	&T_\te(r_h,\t) \bigg[\f{R(r_h,\t)}{R(r,\t)}\bigg]^{\f{1}{6}}\,, \quad &z\gg z_c\\
\end{aligned} \right.
\eea
where $n_\te(r_h,\t)$ and $T_\te(r_h,\t)$ are the boundary values of $n_\te$ and $T_\te$, respectively. Besides, $z_c$ is a characteristic quantity corresponding to $\T_\ion \equiv k_BT_\text{ion}/m_{\text{ion}} = \T_\te$ near the horizon. In the case of $z\ll z_c$, $\T_\ion \ll 1$, indicating that protons are considered to be non-relativistic particles. Conversely, $z\gg z_c$ corresponds to $\T_\ion\gg1$, where protons satisfy the conditions of being ultra-relativistic particles. To facilitate further analysis, we adopt a Gaussian distribution in the $\theta$ direction for $n_\te(r_h, \theta)$ and take $T_\te(r_h, \theta)$ to be a constant. The expressions are as follows:
\bea\label{02}
n_\te(r_h,\t)=n_i \exp \bigg[ -\bigg(\f{\sin{\t}-\sin{\t_J}}{\s}\bigg)^2 \bigg] \, , \quad T_\te(r_h,\t) =T_i \, ,
\eea
Here, $\theta_J$ represents the mean position in the $\theta$ direction, and $\sigma$ characterizes the standard deviation of the distribution. From now on, we set the boundary values of electron temperature and number density as $T_i=10^{11} \, \text{K}$, $n_i = 10^{12} \text{m}^{-3}$, which is applicable to the plasma surrounding the M87* \cite{Vincent:2022fwj, Hou:2023bep}. With such a choice of parameters, we can estimate that $\T_\te\simeq15$ and $z_c \simeq 122$. Furthermore, in practical situations, the temperature of protons can hardly reach the level of being considered highly relativistic particles, and we will focus solely on the case where $z\ll z_c$. 

Regarding the electromagnetic field $F_{\m\n}$ accompanying the fluid, it satisfies $F_{\m\n}u^\n=0$, which implies the absence of the electric field in the rest frame of the fluid. As given in \cite{Hou:2023bep}, the expression for the magnetic field is as follows:
\bea\label{B4}
B^\m= \f{1}{\sqrt{-\text{det}(g_{\m\n})}} \f{F_{\t\p}}{u^r} \big( u_t u^\m +\delta_t^\m \big) \, , \quad  \m=t, r,\t,\p \, ,
\eea
where $\text{det}(g_{\m\n})=-\S^2\sin^2{\t}$ is the determinant of the metric in Kerr spacetime, expressed in BL coordinates, and $F_{\t\phi}$ is a function of $\theta$ for conical motions and can be chosen to be 
\bea\label{psie}
F_{\t\p} = \Psi_0 \, \text{sign}(\cos{\t})  \sin{\t}  \,,
\eea
in subsequent computations, where $\Psi_0$ is a constant indicating the strength of the magnetic field. Furthermore, the other non-zero components of the corresponding electromagnetic field tensor are 
\bea\label{fot}
F_{\p \t}=-F_{\t\p}\,,\quad F_{r\p}=-F_{\p r}=\frac{u^\t}{u^r}F_{\t\p},\,\quad F_{r\t}=-F_{\t r}=\frac{u^\p}{u^r}F_{\t\p}\,.
\eea

\section{Black hole imaging from thermal synchrotron radiation}\label{sec3}

In this section, we briefly review the mechanisms of thermal synchrotron radiation in the fluid medium. The details of radiation process can be found in Appendix \ref{AppA}. Based on this, we  introduce the radiative transfer equation as well as the numerical methods employed in the imaging process. Recall that we have set $T_i=10^{11} \, \text{K}$ in Eq.~\eqref{02}, thus we have 
\bea
\T_\te \equiv \frac{k_B T_\te}{m_\te} \approx 15\,,
\eea
corresponding to highly relativistic electrons in the plasma, where $k_B$ is the Boltzmann constant and $m_\te$ represents the electron's mass. In this regime, the radiation emitted by an electron is predominantly concentrated along the direction of its motion. In our study, we employ the approximated expression for the emissivity of thermal synchrotron radiation presented in \cite{leung2011numerical}, which has the form of
\bea\label{jape}
\mathrm{j}_\n=\frac{2^{-1/2}\pi  e^2}{3} \f{ n_\te \nu}{\T_\te^2}\left(1 + 2^{11/12} X^{-1/3} \right)^2 \exp{ \left[-X^{1/3}\right]}\,, 
\eea
with
\bea
X \equiv \frac{9\pi m_\te \n}{e\T_\te^2\mE}\,, 
\eea
where $\n$ is the frequency of the emitted photons, and 
\bea\label{mee}
\mE = \sqrt{\left(\vec{E}+\vec{n}\times\vec{B}\right)^2-\left(\vec{n}\cdot\vec{E}\right)^2} = \frac{\sqrt{F^{\m\a} F_{\m\b}k_{\a}k^{\b}}}{\nu}  \,
\eea 
represents the strength of the electromagnetic field inducing the curvature radiation of the electrons, where $\vec{n}$ is the unit vector along the radiation direction in the fluid rest frame, and $k^\n$ is the 4-momentum of the emitted photons.  It should be noted that the relative error of $\mathrm{j}_\n$ is within $10\%$ when $\T_\te>1.5$, and within $1\%$ when $\T_\te>5$. Hence, $\T_\te \approx 15$ provides a high accuracy of approximate emissivity given by Eq. (\ref{jape}).

As for the magnetic field, we set a characteristic field strength at $(r_h,\,\t_J)$, that is, $B_i \equiv B(r_h, \t_J)=\sqrt{B^\mu B_\mu}\,\big|_{r=r_h,\, \t=\t_J}$. In the subsequent numerical calculations, we take $B_i = 10$ Gauss, roughly the order of magnitude of the magnetic field outside the M87* \cite{EventHorizonTelescope:2021srq}. As a result, for the conical solution of the fluid, the emissivity can be written as
\bea\label{j3}
\f{ \mathrm{j}_\n}{k_BT_i} = \left(2.94 \times 10^{35} \right)   \bigg[\f{n_i}{10^{12} \text{m}^{-3}}\bigg] \bigg[\f{\n_o}{230\text{GHz}}\bigg] \frac{f(X)}{g}  \bigg[\f{R(r_h,\t)}{R(r,\t)}\bigg]^{\f{2-z}{6(2+z)}} \mathrm{e}^{-\big(\f{\sin{\t}-\sin{\t_J}}{\s}\big)^2}\,,
\eea
at $z\ll z_c$, where $\n_o = \n g$ is the frequency of observed photons, $g$ denotes the redshift factor, and $f(X) \equiv \big(1 + 2^{11/12} X^{-1/3} \big)^2 \mathrm{e}^{-X^{1/3}}$. The dimensionless quantity $X$ is given by
\bea
X  = \left(1.30 \times 10^{2} \right) \bigg[\f{\n_o}{230\text{GHz}}\bigg] \bigg[\f{10\text{Gauss}}{B_i}\bigg] \frac{1}{g^2}\sqrt{\f{(E^2-1)\S(r_h, \t_J)+2r_h}{\S(r_h, \t_J) R(r_h, \t_J) \bar{F}_{\m\a}\bar{F}^{\m\b}\bar{k}^\a \bar{k}_\b}} \bigg[\f{R(r,\t)}{R(r_h,\t)}\bigg]^{\f{2(1+z)}{3(2+z)}}\,, 
\eea
with the renormalized quantities $\bar{k}^{\m} = k^\m / \n_o$, $\bar{F}_{\m\n} = F_{\m\n}/\Psi_0$. 

In order to obtain the image at the observer's location, we need to propagate the radiation from the light source to the observer's screen. Under the geometrical optics approximation, the photons move along null geodesics and carries the information of the medium by accumulating specific intensity. The radiative transport equation for the specific intensity $I_\n$ can be expressed as \cite{Lindquist:1966igj}
\bea\label{RTE}
\f{d}{d\lm} \bigg( \f{I_\n}{\n^3} \bigg)= \f{j_\n - \a_\n I_\n}{\n^2}   \, ,
\eea 
where $\lm$ is the affine parameter along the null geodesic, $\alpha_\n$ is the absorption coefficient of the plasma fluid. If the absorption can be neglected, the scaling of the emissivity results in the observed intensity scaling in the same way, $j_{\n}\rightarrow \epsilon j_{\n}$, $I_{\n_o} \rightarrow \epsilon I_{\n_o}$. However, in the presence of non-zero absorption, such a scale transformation does not exist. In this work, we take the absorption into account, and thus, when handling the quantity $j_\n$, it is essential to retain the constant terms within the expression of $j_\n$. 
If the geodesic of light is solved, which we denote as $x^{\m}(\lm)$, then the solution of Eq.~\eqref{RTE} takes 
\be\label{Inuo}
I_{\n_o} = \n_o \int^{\lm_o}_{\lm_s} d\lm \, g^2\big(x^{\m}(\lm),k^{\m}(\lm)\big)\, j_{\n}\big(x^{\m}(\lm),k^{\m}(\lm)\big) \, \exp\bigg\{-\int^{\lm_o}_{\lm} d\lm' \n\a_{\n}(x^{\m}(\lm'),k^{\m}(\lm'))\bigg\}\, ,
\ee
where $\lm_s$ is the affine parameter of the geodesic integration at the start point, $\lm_o$ is the affine parameter of the light reaching the observer. As indicated by Eq. (\ref{Inuo}), it is necessary to obtain the null geodesic $x^\m(\lambda)$ in order to establish a connection between the light source and the screen. In this study, we adopt the ray-tracing method in \cite{Hu:2020usx} to numerically solve for the null geodesics. After obtaining $x^\m(\lm)$ and $k^\m(\lm) = dx^\m(\lm)/d\lm$, by utilizing Eq. (\ref{Inuo}), we can acquire the intensity of the image on the observer's screen. Additionally, note that in the unit with $GM=c=1$, we still have the freedom to fix the thermal dimension. Since $[I_\n] = [j_\n] = [k_B T_i]$, the observed intensity can be measured in the unit of $k_B T_i$. In the following sections, we set $k_B T_i = 1$ for simplicity.

\section{Results of images}\label{sec4}

In this section, we present the images of the black hole illuminated by the analytical fluid model, based on the methodology outlined in the preceding sections.

\subsection{Images of thick disks}\label{sec33}

This subsection commences with the optical apparance of thick disks surrounding black holes within our theoretical framework. The disk exhibits equatorial symmetry, and the boundary condition of the accretion flow is chosen as
\bea
\t_J = \f{\pi}{2} \, , \quad \sigma_r = -1 \, , \quad E = 1 \,.
\eea
According to Eq. (\ref{LL}), when $E$ equals $1$, $L$ becomes $0$, corresponding to the infalling geodesics. Note that the azimuthal motion does not vanish due to the frame-dragging effect. This resembles the second scenario discussed in \cite{Vincent:2022fwj}, which also considers an infalling accretion flow. However, in contrast to the phenomenological power-law functions of $r$ adopted in \cite{Vincent:2022fwj}, our model derives the distributions of particle number density, temperature, and magnetic field in a self-consistent manner \cite{Hou:2023bep}.

The coefficient $\sigma$ in Eq.~\eqref{02} signifies the thickness of the disk, and we set $\sigma = 0.2$ in the ensuing computations. Moreover, as previously mentioned, we focus on the situation where $z\ll z_c\simeq122$. For the sake of simplicity, we set \footnote{A prevalent formula extensively employed in GRMHD is $T_{\ion}/T_\te = R_{\text{high}}\b^2/(1+\b^2) + 1/(1+ \b^2)$ \cite{Moscibrodzka:2015pda}, where $\b \equiv p_{\text{gas}}/p_{M}$, and $R_{\text{high}}$ is a phenomenological parameter ranging from 1 to 160 for the M87* \cite{EventHorizonTelescope:2019pgp}. Since $p_{M} \approx B^2_i/2 \approx 0.72 \times 10^{52}$, $ p_{\text{gas}} \approx n_ik_BT_i(1+z) \approx 10^{51}(1+z)$. Thus, for $z \approx 1$ we have $\b^2 \ll 1$, and if $z > 6.2$ we may have $\b^2 \gg 1$ and $z \approx R_{\text{high}}$. Therefore, we can choose $z = 1$ for jets and $z=20$ for disks in our model.} $z=20$. Additionally, we have set the characteristic magnetic field strength $B_i=10$ Gauss for our subsequent research, a choice informed by the order of magnitude of the magnetic field outside the M87*.

\begin{figure}[h!]
\centering
\includegraphics[width=6.8in]{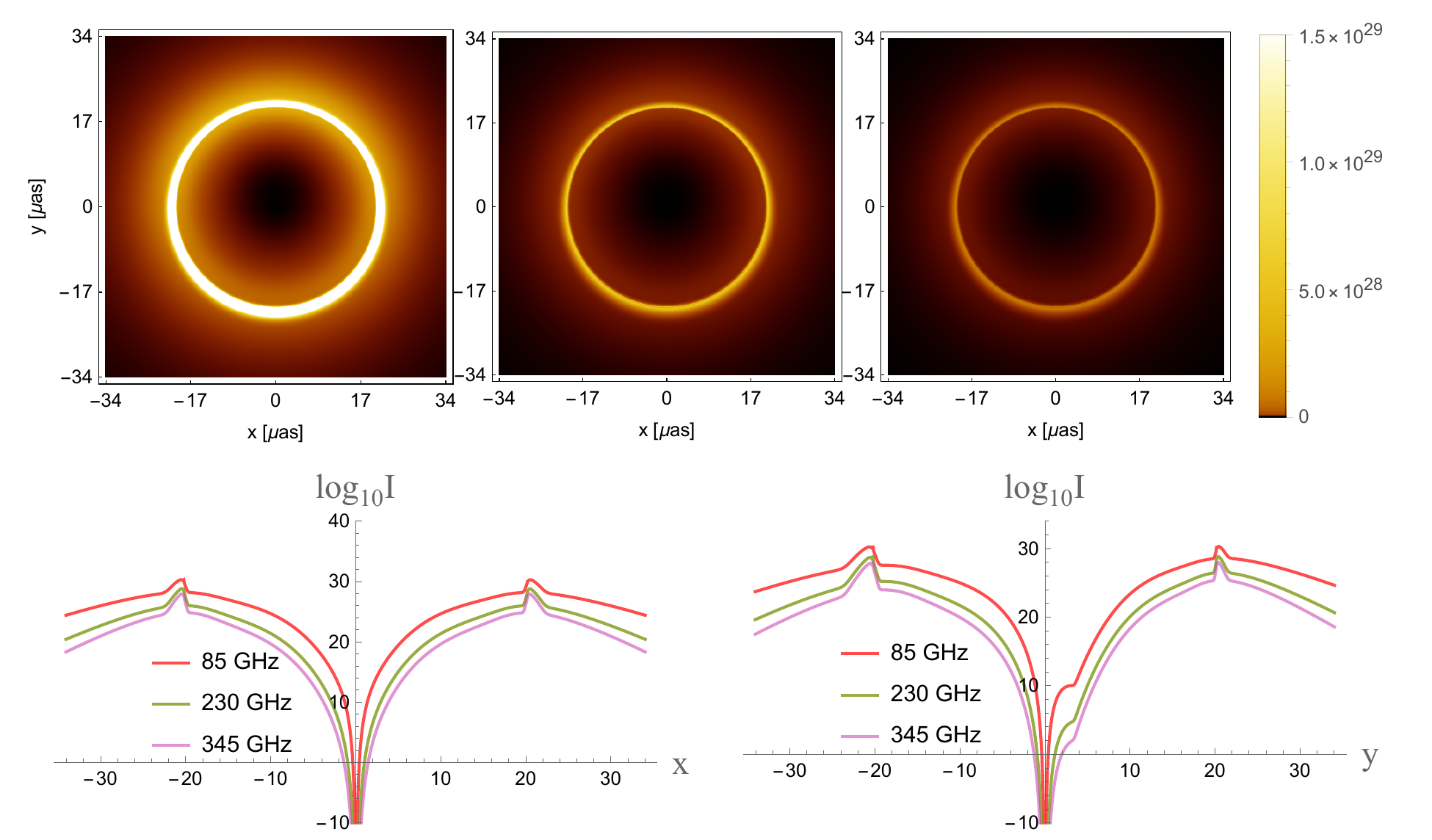}
\caption{Upper row: intensity maps depicting a black hole with $a=0.01$, illuminated by a thick disk, observed at an inclination of $17^\circ$. From left to right, the observational frequencies are $85$ GHz, $230$ GHz, and $345$ GHz. Lower row: the horizontal (left) and vertical (right) intensity cuts across various frequencies.}
\label{fig:17001}
\end{figure}
	 
\begin{figure}[h!]
\centering
\includegraphics[width=6.8in]{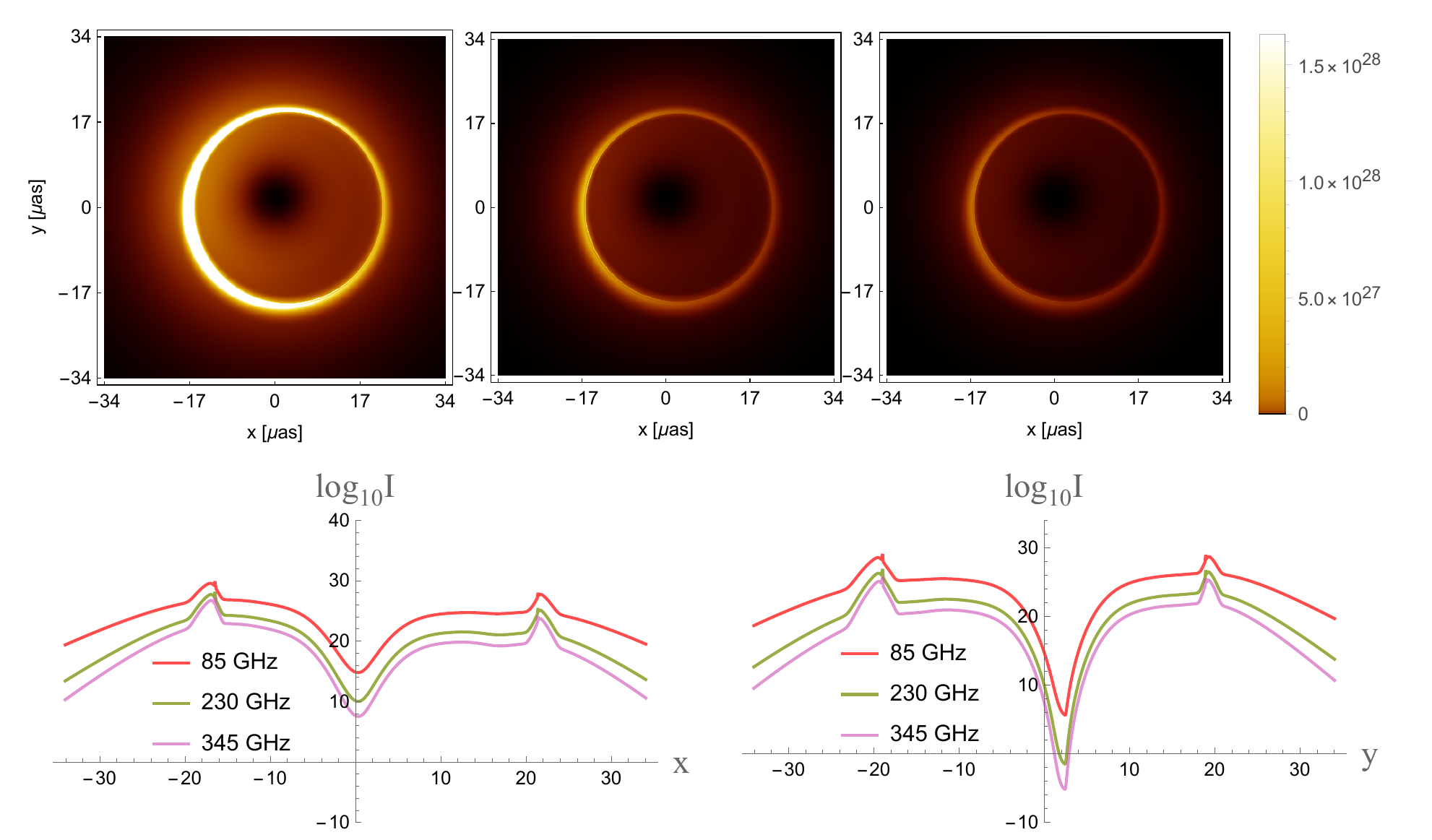}
\caption{Upper row: Intensity maps depicting a black hole with a spin parameter $a=0.94$. This black hole is illuminated by a thick disk and is observed at an inclination of $17^\circ$. The maps are arranged from left to right according to their observation frequencies, which are $85$ GHz, $230$ GHz, and $345$ GHz respectively. Lower row: the horizontal (left) and vertical (right) intensity cuts for different frequencies.}
\label{fig:17094}
\end{figure}

\begin{figure}[h!]
\centering	
\includegraphics[width=6.8in]{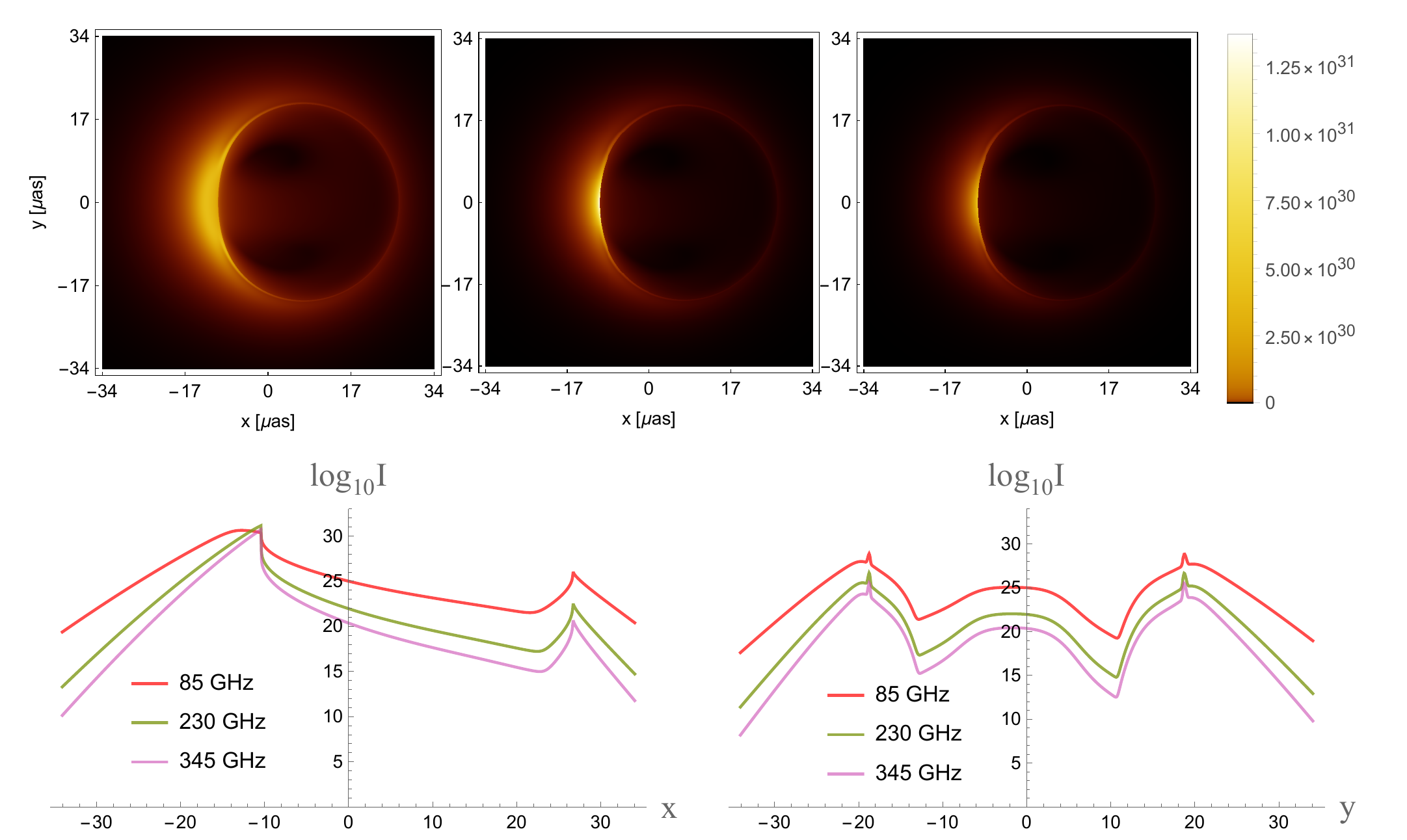}
\caption{Upper row: intensity maps depicting a black hole with a spin parameter $a=0.94$. This black hole is illuminated by a thick disk and is observed at an inclination of $80^\circ$. The maps are arranged from left to right according to their observation frequencies, which are $85$ GHz, $230$ GHz, and $345$ GHz respectively. Lower row: the horizontal (left) and vertical (right) intensity cuts for different frequencies.}
\label{fig:80094}
\end{figure}

In Figs.~\ref{fig:17001} and \ref{fig:17094}, we present the imaging results for geometrically thick accretion disks surrounding a black hole with spin parameters $a=0.01$ and $a=0.94$ respectively. These images are viewed from an observational angle of $\theta_o=17^\circ$. In Fig.~\ref{fig:80094}, we also provide the imaging results at the observational angle $\theta_o=80^\circ$ with $a=0.94$. The upper row of each figure illustrates the light intensity distribution of the black hole images. These images are arranged from left to right according to their observation frequencies, which are $85$ GHz, $230$ GHz, and $345$ GHz respectively. The lower row, also ordered from left to right, depicts the light intensity distribution along the lines where $y=0$ (horizontal cuts) and $x=0$ (vertical cuts). Specifically, for the light intensity cuts displayed in the lower row, we have taken the common logarithm of the light intensity in order to see clearly the change in light intensity. 

Within all the intensity cuts diagrams, it can be discerned that each line is characterized by two prominent peaks. The regions outside these peaks correspond to the primary image of the accretion disk, with light from these regions reaching the observer directly from the disk. Conversely, the peaks themselves represent higher-order images, where the light has circumnavigated the black hole once or multiple times prior to reaching the observer. These higher-order images manifest as conspicuous rings within the intensity maps. These rings are a consequence of the strong gravitational lensing effect, whereby the gravitational attraction of the black hole warps the path of light, enabling a portion of it to bypass the black hole and eventually reach the observer. From the three illustrations, it can be observed that the ring structure at $85$ GHz is significantly brighter than those at $230$ GHz and $345$ GHz. The luminosity of the accretion disk image exhibits a pronounced dependency on the observation frequency; as the observation frequency increases, the intensity diminishes. This phenomenon is more conspicuous when the spin is greater, see Fig. \ref{fig:17094} for  $a=0.94$. However, the position of the higher-order images remains unchanged across three different frequencies, as the gravitational lensing effect exerted by the black hole is independent of the light's frequency. Moreover, in every image featuring a thick disk, the ring structure is distinctly visible. The lensing effect enhances the light intensity near the location of the photon ring by 3-5 orders of magnitude. This significant enhancement makes the ring structure highly prominent in the disk model. 
	 
Beyond the ring structure, another noteworthy feature is the darkest region observable from various angles. That region originates from the event horizon itself. In the case of a geometrically thin disk extending to the horizon, the primary image of the horizon, also known as the ``inner shadow'', has a distinct outline and might be captured by the EHT \cite{Chael:2021rjo}. However, for a geometrically thick disk, the horizon's outline is obscured by the emission outside the equatorial plane, making it more challenging to observe. Despite this, an extremely dark region is still visible to an observer at $\t_o=17^\circ$, as illustrated in Figs.~\ref{fig:17001} and \ref{fig:17094}. Furthermore, in Fig. \ref{fig:80094} where $\t_o=80^\circ$, two distinct black areas emerge, with the upper one being slightly darker than the lower one, a phenomenon attributed to the lensing effect.

\begin{figure}[h!]
\centering	
\includegraphics[width=6.8in]{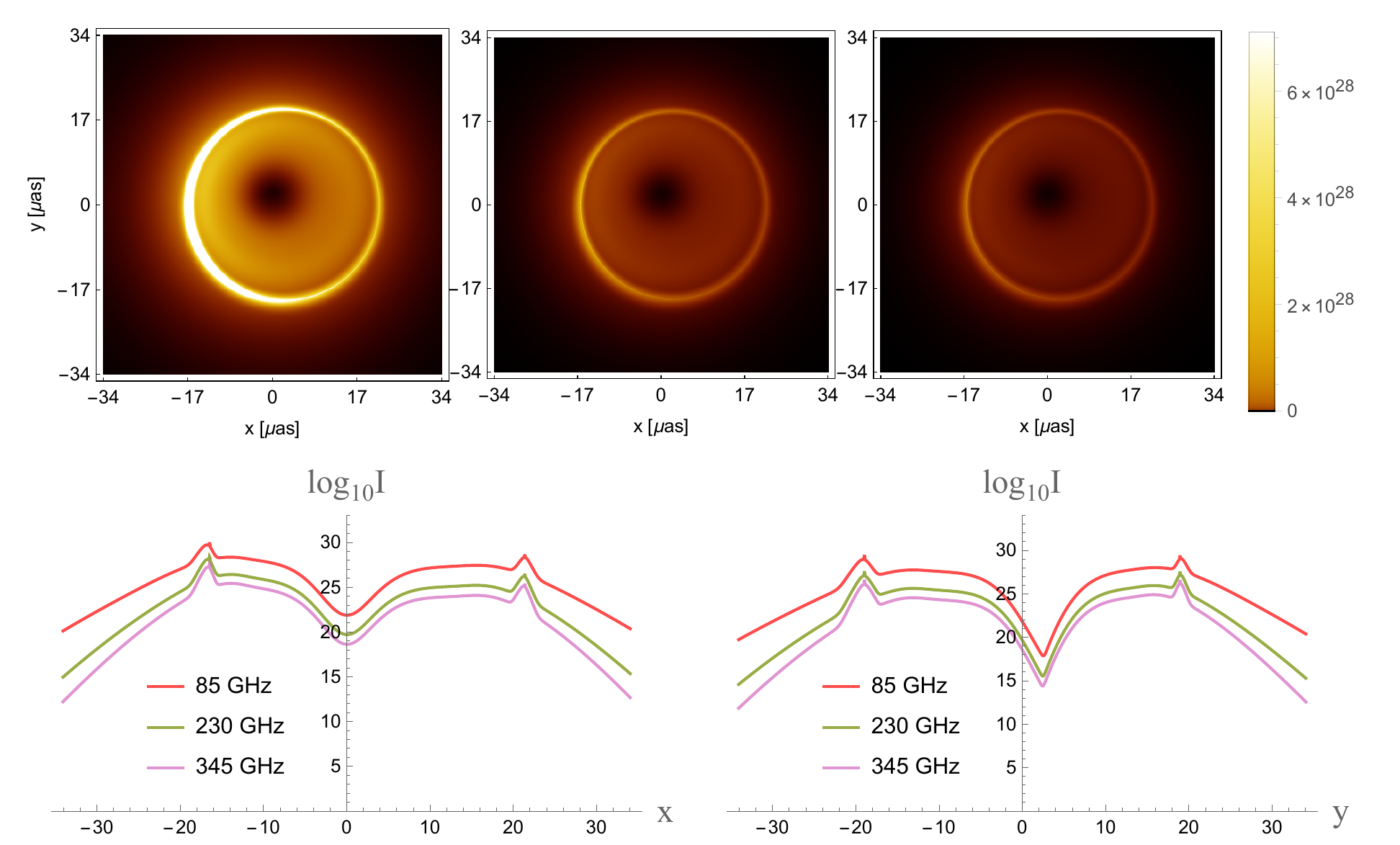}
\caption{Upper row: the isotropic intensity maps representing a black hole with a spin parameter of $a=0.94$. This black hole is illuminated by a thick disk and observed at an inclination of $17^\circ$. The maps are arranged from left to right according to their observation frequencies, which are $85$ GHz, $230$ GHz, and $345$ GHz respectively. Lower row: the horizontal (left) and vertical (right) intensity cuts for different frequencies.}
\label{fig:avrgi}
\end{figure}

\begin{figure}[h!]
\centering	
\includegraphics[width=6.9in]{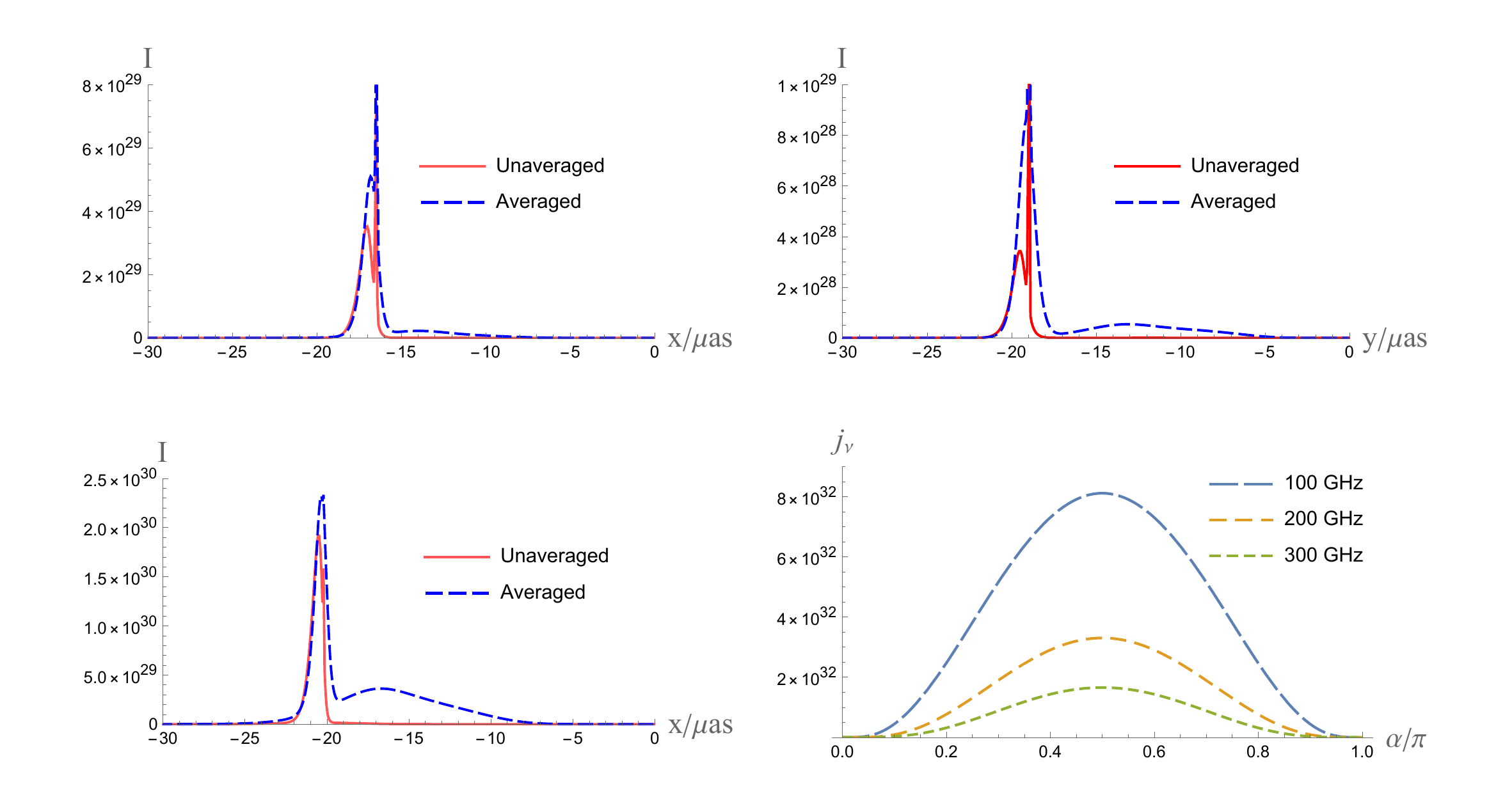}
\caption{Upper row: A comparison of the intensities of $\n_o = 85$ GHz generated by both the averaged and the unaveraged emissivity along the horizontal axis (left) and the vertical axis (right) at $a=0.94$. Lower row: A comparison of the intensities of $\n_o = 85$ GHz generated by the averaged and the unaveraged emissivity at $a=0.01$, as well as the anisotropic emissivity as a function of the pitch angle. The emission position is given by $r=r_h +0.5$, $\t = \pi/2$. The observation angles for both the upper and lower rows are $\theta_o=17^\circ$.}
\label{fig:Avrg}
\end{figure}

It is particularly important to emphasize that synchrotron radiation is highly anisotropic, which means that the radiative power varies with the direction of emission, represented by the pitch angle $\a$ which represents the angle between the wave vector and the magnetic field vector in the fluid rest frames. Specifically, we can rewrite  Eq.\eqref{mee} as $\mE=\big|\vec{n}\times \vec{B}\big| =  \big|\vec{B}\big| \sin{\a}$ in the ideal MHD fluid. To study the influence of such anisotropy on the black hole images, it is necessary to consider an isotropic emission profile as a comparison. Here, we employ an angle-averaged emissivity defined as
\be\label{JA}
\bar{j}_\n =\frac{1}{4\pi}\int\mathrm{j}_\n d\Omega= \int^{\pi/2}_{0} \mathrm{j}_\n \sin{\a} d\a \, ,
\ee
where $d\Omega=\sin\a d\a d\zeta$ is the solid angle element and $\zeta$ denotes the azimuthal angle in the fluid rest frames. We aim to compare the results generated by Eq.~\eqref{jape} and Eq.~\eqref{JA}. In Fig. \ref{fig:avrgi}, we present the images generated by $\bar{j}_\n$ , where the other parameters are kept consistent with those in Fig. \ref{fig:17094}. The distinctions are noticeably clear. The most conspicuous difference is that at the same frequency, the brightness of the primary image in Fig. \ref{fig:avrgi} is greater than that in Fig. \ref{fig:17094}. This difference can be more clearly observed from the intensity cuts. Regardless of whether the intensity cuts are along the horizontal or vertical direction, the changing trend in the primary image of Fig. \ref{fig:17094} is obviously steeper than the one in Fig. \ref{fig:avrgi}.

In Fig. \ref{fig:Avrg}, we present a comparative chart of intensity cuts from isotropic and anisotropic emissions observed at $\theta_o=17^\circ$. To see the details of the peaks clearly, we restrict to the sections of the intensity cuts with either $x<0$ or $y<0$. The first row provides a comparison of the intensities at $\n_o = 85$ GHz along the horizontal axis (left) and the vertical axis (right) at $a=0.94$. The second row presents the case for $a=0.01$. The left panel displays the comparison of intensities at $\n_o = 85$ GHz, generated by both the averaged and unaveraged emission along the horizontal direction. Given that the black hole is almost non-rotating at $a=0.01$, the horizontal and vertical directions are nearly identical, so we only plot the horizontal intensity cut. Additionally, the right panel illustrates the emissivity $\mathrm{j}_\n$ as a function of the pitch angle $\a$. The emissivity peaks when $\a=\pi/2$, where the radiation direction is perpendicular to the magnetic field. This fact remains valid even in the high-spin case, as the emissivity is insensitive to the spin of the black hole. Notably, the emissivity exhibits a significant increase as the frequency decreases. 

According to Fig. \ref{fig:Avrg}, it becomes apparent that the primary image generated by the isotropic emission is brighter within the peaks compared to the anisotropic scenario. Such a distinction is closely related to the magnetic field configuration. To see this point more clearly, let us consider a simple case of a Schwarzschild black hole observed almost face-on, with the disk residing on the equatorial plane, and a purely radial magnetic field described by Eq.~\eqref{B4}. Gravitational lensing imparts a significant radial component to the three-momentum of the photon emitted from the disk near the horizon and reaching the observer. Consequently, the emission direction and magnetic field line maintain a small angle in the emission region.  This results in the suppression of synchrotron emission in the presence of a radial magnetic field, as shown in the bottom left panel of Fig. \ref{fig:Avrg}. This suppression becomes especially pronounced near the horizon, where the lensing effect intensifies. In contrast, the photons corresponding to the lensed images form the photon sphere at a constant radius. Their three-momenta are nearly perpendicular to the magnetic field in the emission region such that the intensity of lensed images is less suppressed compared to that of primary images.
 
\begin{figure}[h!]
\centering
\includegraphics[width=6.4in]{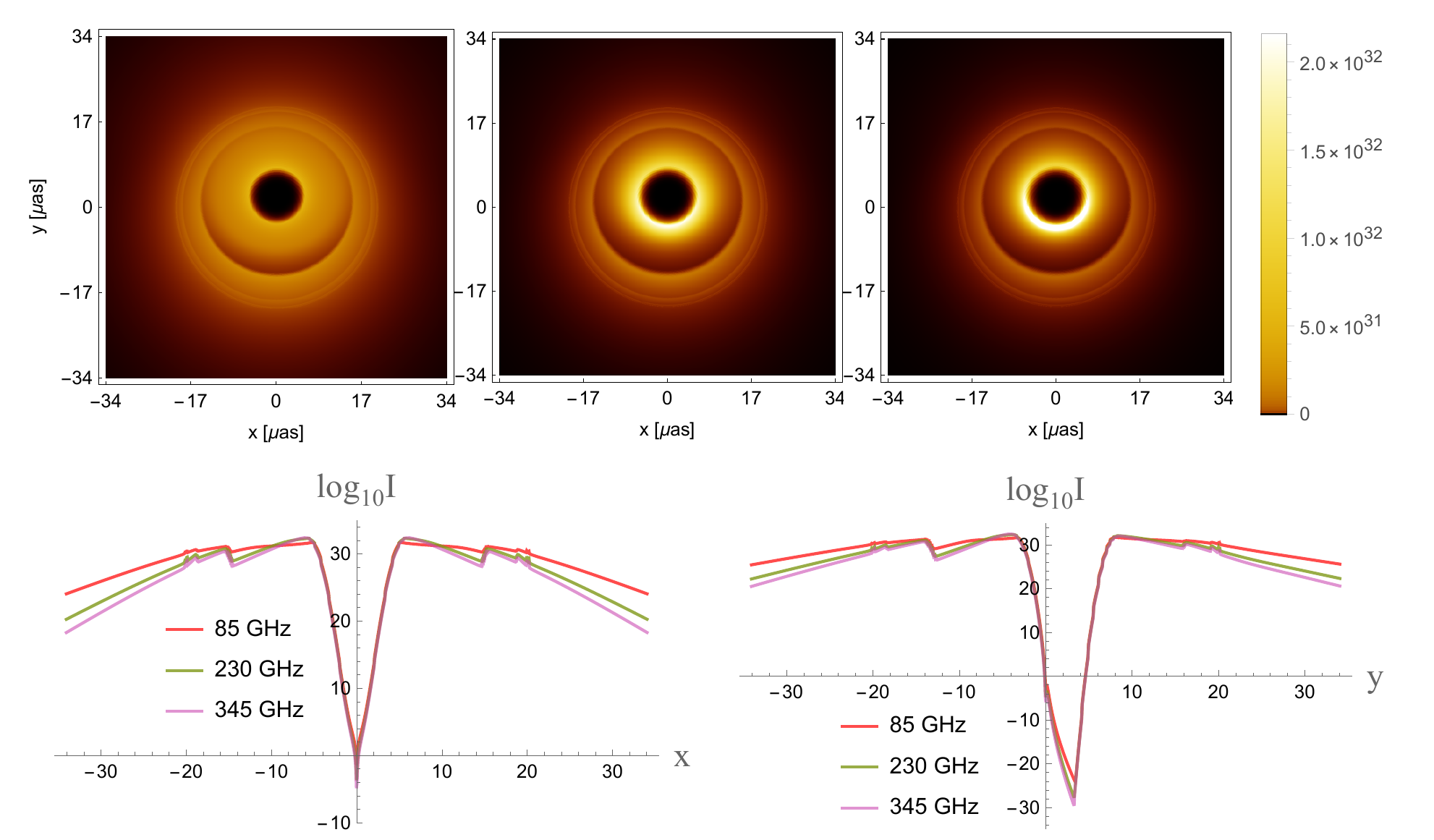}\\
\includegraphics[width=6.4in]{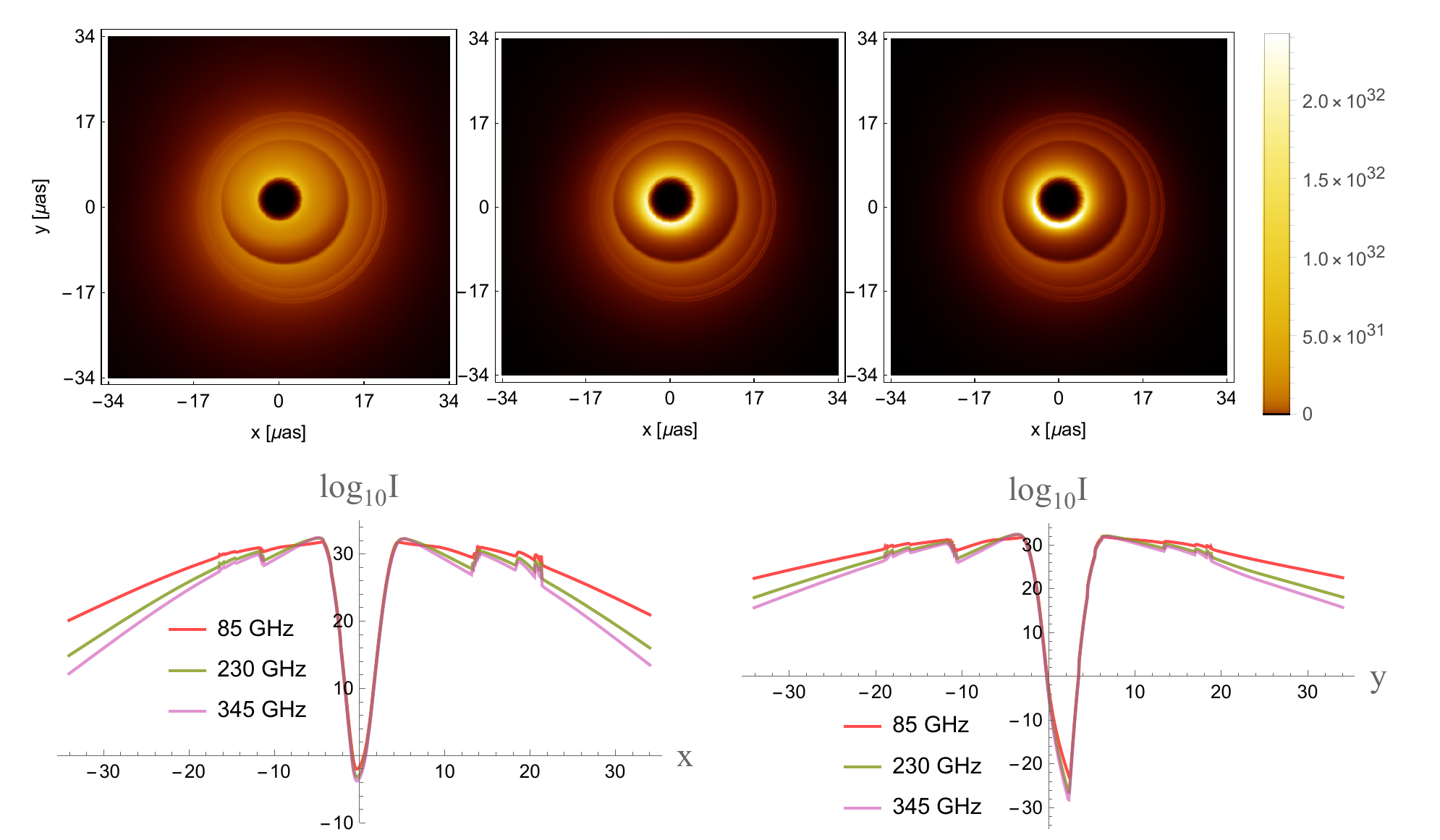}
\caption{The first and third rows represent the intensity maps of a black hole with $a=0.01$ and $a=0.94$ respectively, illuminated by an outflow jet and observed at an angle of $\t_o = 17^{\circ}$. From left to right, the observation frequencies are $85$ GHz, $230$ GHz and $345$ GHz. The second and fourth rows correspond to the horizontal (left) and vertical (right) intensity cuts at different frequencies for the situations depicted in the first and third rows, respectively.}
\label{out17}
\end{figure}

\subsection{Images produced by a funnel wall}\label{sec32}

In this subsection, we explore the features of black hole images illuminated by the funnel wall in our model. The shape of the funnel wall is obtained by setting
\bea
\t_J = \f{\pi}{4} \, , \quad  \s=0.05 \, , 
\eea	
with $\t_J$ denotes the open angle of the funnel, $\sigma$ is small and represents a geometrically thin wall. In the funnel wall region, we opt for $z=1$, a choice validated by the GRMHD simulations in \cite{EventHorizonTelescope:2019pgp}. Furthermore, since both inward and outward flows are observed in the simulations, we consider both scenarios, represented by $\sigma_r = \pm$ respectively. In the case of the conical model, there remains a free parameter $E$ denoting the velocity of the flow. In this study, we select $E=1$ to represent a mild flow. As for the magnetic field, we persist in choosing  a characteristic strength of $B_i=10$ Gauss.

\begin{figure}[h!]
\centering
\includegraphics[width=6.8in]{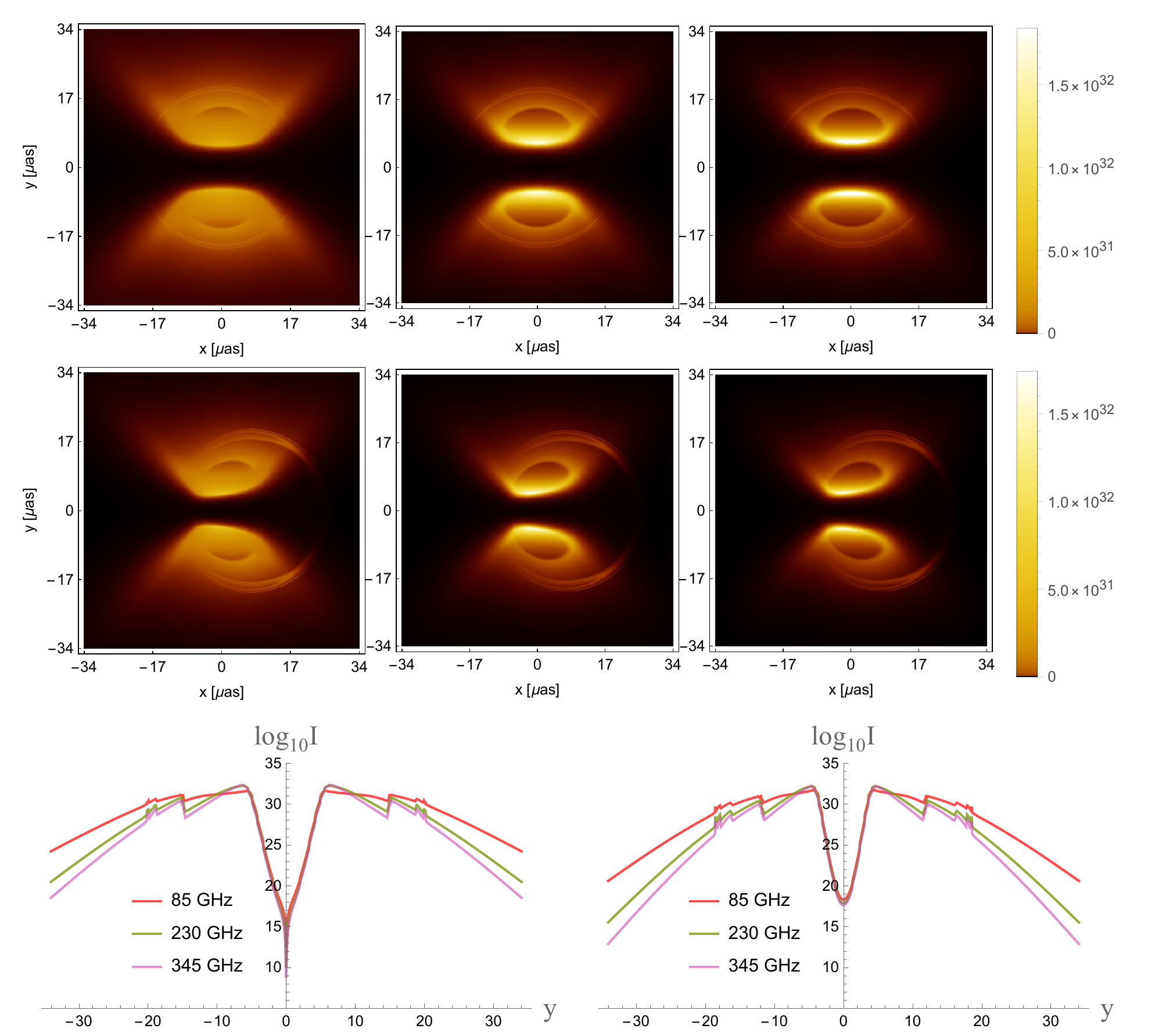}
\caption{The first two rows represent the intensity maps of a black hole with $a=0.01$ and $a=0.94$ respectively, illuminated by an outflow jet and observed at an angle of $\t_o = 90^{\circ}$. From left to right, the observation frequencies are $85$ GHz, $230$ GHz and $345$ GHz. The left and right images in the third row respectively represent vertical intensity cuts at different frequencies for the scenarios illustrated in the first twoh rows.}
\label{out90}
\end{figure}

\begin{figure}[h!]
\centering
\includegraphics[width=6.8in]{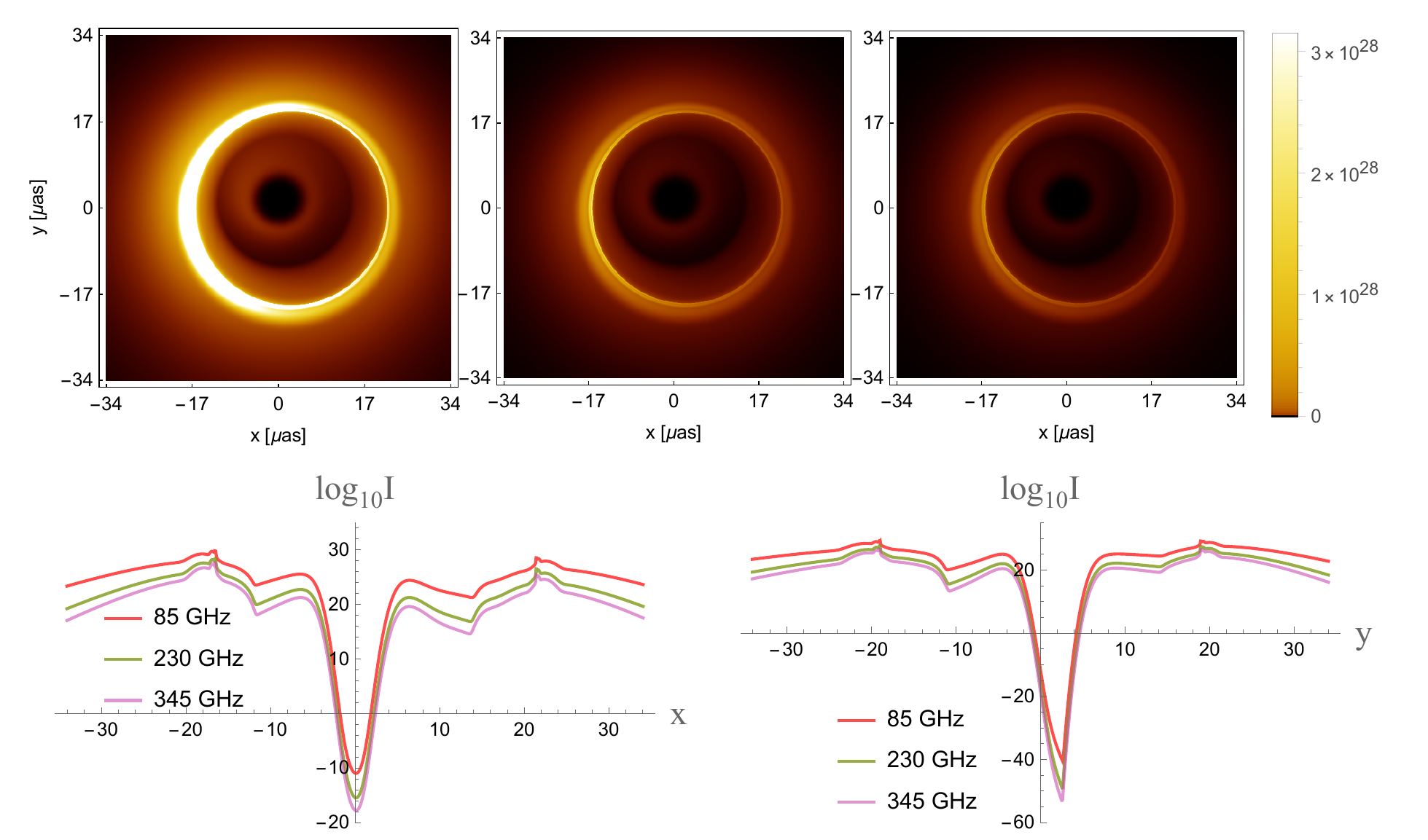}
\caption{ Upper row: intensity maps of a Kerr black hole with $a=0.94$ illuminated by an inflow funnel wall observed at $\t_o = 17^{\circ}$. From left to right, the observation frequency are $85$ GHz, $230$ GHz, $345$ GHz. Bottom row: the horizontal (Left) and vertical (Right) intensity cuts of different frequencies.}
\label{fig:IFJ}
\end{figure}

\begin{figure}[h!]
\centering
\includegraphics[width=6.8in]{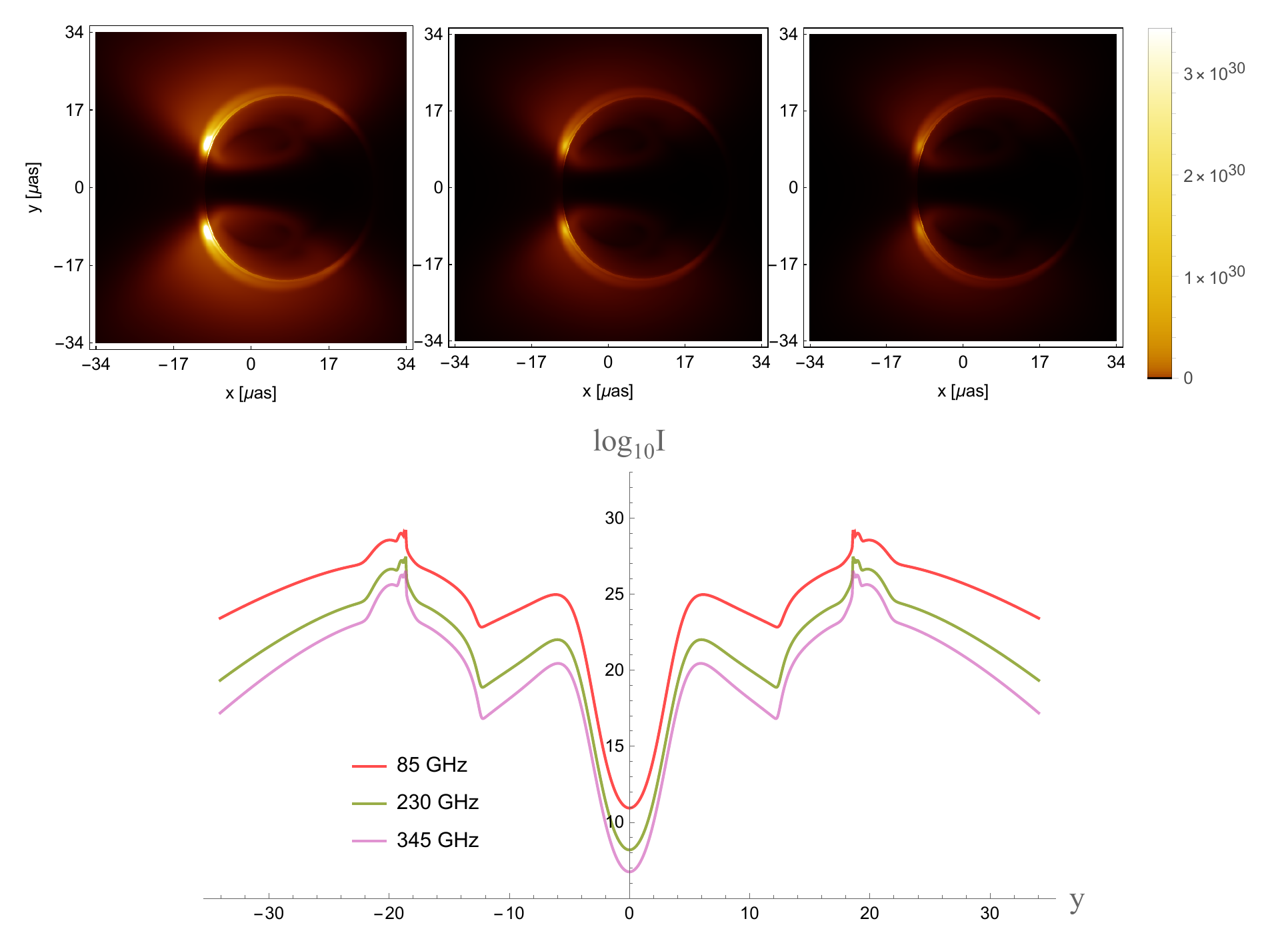}
\caption{ Upper row: intensity maps of a Kerr black hole with $a=0.94$ illuminated by an inflow funnel wall observed at $\t_o = 90^{\circ}$. From left to right, the observation frequency are $85$ GHz, $230$ GHz, $345$ GHz. Bottom row: the vertical intensity cuts of different frequencies. We neglect the horizontal intensity cuts as they are almost invisible.}
\label{fig:IFJth90}
\end{figure}

The imaging results of the outflow are showcased in Figs. \ref{out17} and \ref{out90}, representing the images at $\t_o=17^\circ$ and $\t_o=90^\circ$ respectively. From the figures, we notice a prominent feature: the high brightness of the primary image of the jet base, attributed to the significant Doppler effect associated with the outflow. In this discussion, we use the term `` jet base '' to refer to the inner boundary of the jet, that is, the intersection of the funnel wall and the event horizon. When observed at $\t_o=17^\circ$, the primary image of the jet base appears like a donut because the observer is situated within the jet cone, $\t_o < \t_J$. However, at $\t_o=90^\circ$, since the observer is outside the jet cone and positioned on the equatorial plane, two symmetrically separated images of the jet base can be seen. Both of these images exhibit increasing brightness and thickness with the increase in frequency. Besides, at $\t_o=17^\circ$, we notice a series of annular structures. This phenomenon can be attributed to two factors. Firstly, the funnel wall manifests as a biconical structure. Secondly, the gravitational lensing effect results in a series of lensed images of the biconical structure. Similar observations can be made when the observational angle is $\t_o=90^\circ$. At this particular angle, symmetrical lensed images become visible.

\begin{figure}[h!]
\centering
{\includegraphics[scale=0.5]{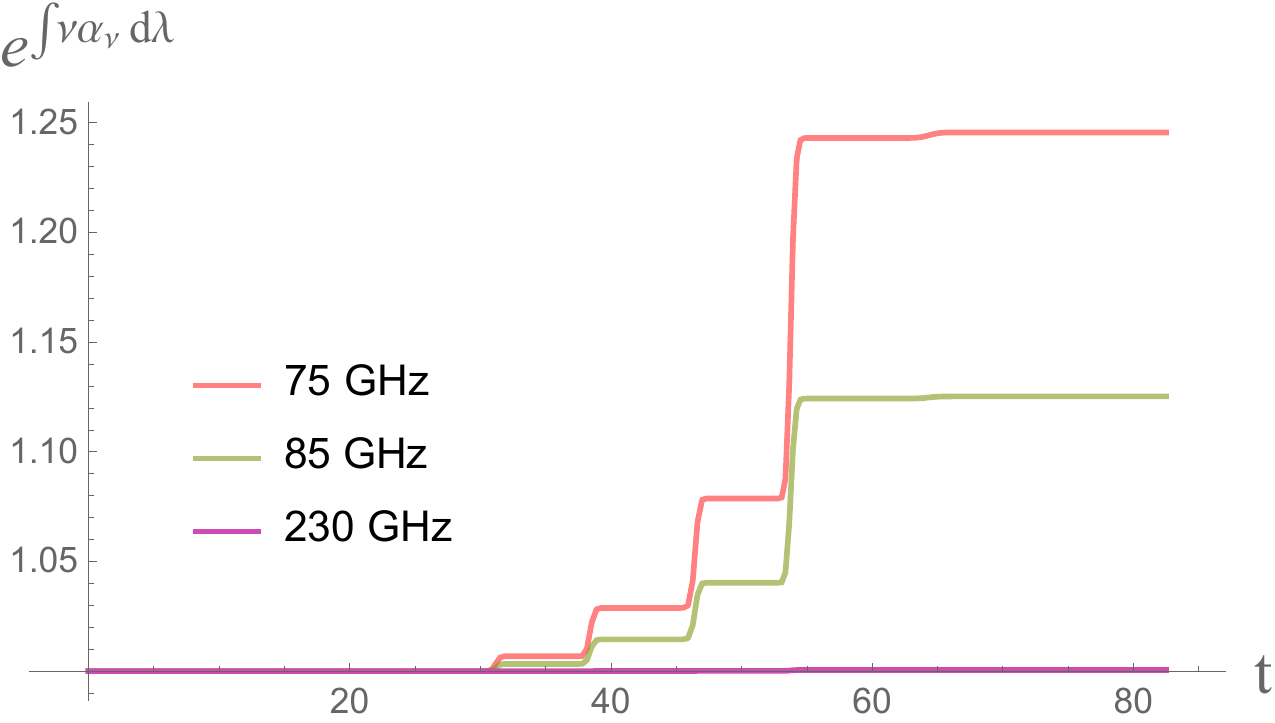}} 
\caption{Optical depth as a function of $t$ along a light ray that crosses the equatorial plane three times. The source is set as a outflow funnel wall, with $a=0.94$ and $\theta_o=17^\circ$. The step increase in the optical depth curves indicates that the light traverses the jet region multiple times.}
\label{fig:tau}
\end{figure}

However, if the emissions near the M87* observed by the EHT originate from the funnel wall, the jet base should not exhibit such high brightness as depicted in Fig. \ref{out17}. This suggests that the outflow model might not be applicable to the near-horizon region of M87*. Additionally, according to the study of GRMHD, in the region below the so-called stagnation surface, the particles are thought to fall to the BH by gravitation \cite{takahashi1990magnetohydrodynamic, Pu:2015rja, Kawashima:2020rmr, Shen:2023nij}. In order to closely match the observed images, we have considered the imaging results of an inflow funnel wall in Figs. \ref{fig:IFJ} and \ref{fig:IFJth90}. Both figures present results for $a=0.94$. The former is observed at $\t=17^\circ$, while the latter is viewed at $\t=90^\circ$. 

In Figure \ref{fig:IFJ}, the primary and lensed images of jet base are clearly visible, along with the distinctive structure of the photon ring, which exhibits the highest brightness. Interestingly, the brightness first decreases as we move inward from the photon ring, then increases, before ultimately drops significantly, a pattern attributable to the primary image of the jet base. In Figure \ref{fig:IFJth90}, where the observer is located on the equatorial plane, the full photon ring is obscured, yet a segment of the ring remains visible. These two figures suggest that the brightness of the primary image of the jet base is significantly diminished compared to the photon ring due to the inward flow. Consequently, the emissions from the inflow funnel wall are not ruled out by EHT's observation. The ring-like image in M87* may contain the important feature of the jet bases in addition to the photon ring \cite{Kawashima:2020rmr}. Our imaging results are also in alignment with the inflow dual cone model presented in \cite{Papoutsis:2022kzp}.

Furthermore, in Figs. \ref{out17} and \ref{out90}, we notice that the intensity cuts of 85 GHz appear less inclined compared to those at 230 GHz and 345 GHz. As manifested in the images themselves, the brightness distribution at 85 GHz seems more smoothed out. This phenomenon may be attributed to the effect of absorption, prompting us to investigate the optical depth along a light ray. The results are depicted in Fig. \ref{fig:tau}, where we concentrate on a specific light ray that starts at $r=15$, initially heads towards the black hole, then deflects and escapes to the observer. The data reveals that the influence of optical depth on radiative transfer significantly varies with frequency. To be specific, the absorption effect is negligible at $230$ GHz, but at lower frequencies such as $85$ GHz or $75$ GHz, it begins to play a significant role. Concurrently, the emissivity is frequency-dependent as well, with a tendency to increase at lower frequencies.  As a result, despite heightened absorption, the observed intensity at lower frequencies remains brighter. It is essential to recognize that the above finding hinges on the positioning of the peaks within the thermal synchrotron radiation spectrum, which in turn depends on the parameters in the model, i.e., $n_i$, $T_i$, and $B_i$. Therefore, it is necessary to determine the physical parameters more accurately in our model based on both the existing and future observational results of M87*.

\section{Summary and discussion}\label{sec5}

In this study, we continued our recent work \cite{Hou:2023bep} and utilized the analytical results of the conical solution for the magnetofluid to investigate black hole imaging from a thick disk and a funnel wall. In our study, the spatial distributions of number density and temperature are explicitly provided by Eq.~\eqref{nT11} and Eq.~\eqref{nT22}, respectively, and the magnetic field structure, exhibiting a highly radial pattern for the conical solution, is described by Eq.~\eqref{B4}. We considered the synchrotron radiation of thermal electrons in the magnetofluid, which was subsequently transferred to a distant observer to generate the images. In practice, we  numerically solved the null geodesic and radiation transfer equations to obtain the images.

We investigated thick disk images at varying observational angles and frequencies. Particularly, to examine the impact of emission anisotropy, we compared the imaging outcomes from normal synchrotron radiation with those from isotropic (angle-averaged) radiation. We noticed that emission anisotropy produces more pronounced photon rings at an observational angle of $\t_o =17^{\circ}$, and in contrast, isotropic emission yields more distinct primary images. This suggests that for synchrotron radiations within a radial magnetic field surrounding a black hole, the photon ring structure is more readily observable for an observer nearly face-on. This insight could be validated by advancements in the next-generation EHT \cite{Ayzenberg:2023hfw}. 

Furthermore, we explored the images of a funnel wall, considering both outward and inward flows. Our study shows that the outflow models yield excessively bright primary images of jet bases, potentially resulting in the absence of the black hole shadow after blurring, inconsistent with current observations of M87*. On the contrary, the images illuminated by the emissions  from the inward funnel wall are still consistent with the observations.

Building upon our findings, we further affirm the validity of our analytical model in characterizing the morphology of thick disks and funnel walls at the horizon scale. However, the model still exhibits certain limitations. Firstly, the model presumes that the magnetofluid is governed by gravity at the horizon scale, which restricts the permissible parameter space. For instance, to fully comprehend the dynamics of funnel wall flows, it is crucial to integrate viscosity into the relativistic Euler equation, a task that presents considerable challenges. Secondly, our analytical results are based on a stationary and axisymmetric fluid configuration, which may not accurately reflect the myriad of astronomical processes that happen in accretion disks or funnel wall regions, such as turbulences, magnetic reconnections, and flares. To procure time-varying results, the application of numerical methods becomes essential. Thirdly, our model operates under the assumption that the electrons form a thermal distribution, with the temperature fraction $z =T_{\text{ion}}/T_{\text{e}}$ being a constant. This assumption may not hold true, particularly within the jet region. Lastly,  the conical flow could be a little too idealized situation, and  it may not capture aspects of even a time-averaged flow.  Nevertheless, we still expect that it provides a satisfactory description of fluid morphology in the near-horizon region of black hole.

\section*{Acknowledgments}
The work is partly supported by NSFC Grant No. 12275004, 12205013 and 11873044. MG is also endorsed by ”the Fundamental Research Funds for the Central Universities” with Grant No. 2021NTST13.

\appendix

\section{{Thermal synchrotron radiation}}\label{AppA}

Electrons in a hot plasma medium can emit thermal synchrotron radiation when they undergo acceleration by electromagnetic fields. It is helpful for us to start with the general form of the power spectrum of thermal synchrotron radiations in the fluid rest frames. For a single electron, the power spectrum of synchrotron radiation is \cite{1954The, 1979Lightman}
\be
P_{\n} = \sqrt{3} \f{e^3}{m_\te} \mE_{\EM} F\bigg(\f{\n}{\n_c}\bigg)\, , \quad  F(x) = x \int_x^{\infty}K_{5/3}(x')dx' \, ,
\ee
where $e$ is the charge of an electron, $m_\te$ represents the mass of an electron, $\n$ denotes the radiation frequency, and $\n_c \equiv 3e\g^2\mE_{\EM}/(2\pi m_\te)$ corresponds to the characteristic frequency. Here, $\g=1/\sqrt{1-v^2}$ is the Lorentz factor and $\mE_{\EM} = \sqrt{\left(\vec{E}+\vec{v}\times\vec{B}\right)^2-\left(\vec{v}\cdot\vec{E}\right)^2}$ encodes the curvature of the electron's orbit caused by the combination of electric field $\vec{E}$, magnetic field $\vec{B}$, and the velocity vector $\vec{v}$ of the moving electron. $K_{5/3}(x^\prime)$ represents the modified Bessel function, which can be written as
\bea
K_{5/3}(x^\prime)=\int_0^\infty e^{-x^\prime\cosh x} \cosh\left(5/3 x\right)dx\,. 
\eea
In our study, we focus on relativistic electrons with high temperature, satisfying $\T_\te \equiv k_B T_\te/m_\te \gg1$. In this regime, the radiations emitted by electrons are  predominantly concentrated along the directions of their motions. Therefore, the emissivity of a single electron can be approximated as
\bea
J_\te \approx P_{\n} \delta\left(\Omega - \Omega_{\vec{v}}\right)\,,
\eea
where $\Omega$ denotes the solid angle along the direction of radiation, $\Omega_{\vec{v}}$ represents the solid angle along the direction of motion and $\delta(x)$ refers to the delta function. Taking the thermal distribution into account, the emissivity of the collection of thermal electrons is given by
\bea\label{j1}
j_{\n} = \f{1}{4\pi} \int d\g d\Omega_{\vec{v}}  \f{dn_\te}{d\g} J_e  \approx \f{\sqrt{3}e^3n_\te}{8\pi\T_\te^3 m_\te}  \mE\int_0^{\infty} d\g \g^2 e^{-\f{\g}{\T_\te}} F\bigg(\f{4\pi m_\te\n}{3e\g^2\mE}\bigg) \, ,
\eea
where
\bea\label{meeApp}
\mE = \sqrt{\left(\vec{E}+\vec{n}\times\vec{B}\right)^2-\left(\vec{n}\cdot\vec{E}\right)^2}\,,
\eea 
with $\vec{n}$ represent the unit vector along the radiation direction, and $n_\te$ denotes the number density of the electrons. Note that we have used $|\vec{v}| \approx 1$, $\T_\te \gg 1$ when deriving the relation \eqref{j1}, since the emission is mainly contributed by ultra-relativistic electrons. With the expansion of $F(x)$, we can obtain two asymptotic expressions for the integral \cite{leung2011numerical},
\bea\label{jap}
j_{\n} \approx
\begin{cases}
\, \f{2^{4/3} \pi e^2}{3}\f{ n_\te  \nu}{\T_\te^2}X^{-2/3} ,& \quad \text{if}\,\,\, \ X \ll 1 \, ;  \\
\,  \\
\, \f{2^{-1/2}\pi  e^2}{3} \f{ n_\te \nu}{\T_\te^2} \exp{ \left[-X^{1/3}\right]}, & \quad \text{if}\,\,\, \ X \gg 1  \, ,
\end{cases}
\eea
where 
\bea
X \equiv \frac{9\pi m_\te \n}{e\T_\te^2\mE}.
\eea
Nevertheless, the expression for $j_\n$ given by Eq. (\ref{jap}) is incomplete, providing only the asymptotic expressions for the cases when $X$ is much greater than 1 and much less than 1. To obtain the complete expression for $j_\n$, we employ the approximation method proposed in \cite{leung2011numerical}, which involves connecting the two asymptotic expressions from Eq. (\ref{jap}) using a fitting function. The expression of the approximate fitting function is given by
\bea\label{japeApp}
\mathrm{j}_\n=\frac{2^{-1/2}\pi  e^2}{3} \f{ n_\te \nu}{\T_\te^2}\left(1 + 2^{11/12} X^{-1/3} \right)^2 \exp{ \left[-X^{1/3}\right]}\,.
\eea

For the convenience of numerical calculations, we would like to rewrite $\mE$ in a covariant form. Note that in Eqs. (\ref{meeApp}) the vectors $\vec{n}$ and $\vec{B}$ are evaluated at the tetrad of the local rest frame of the fluid $\{e^{\m}_{(0)} = u^\m, e^{\m}_{(i)},\, i=1, 2, 3\}$. We set $k^{\m}$ to be the 4-momentum of the emitted photons, and the 3-dimensional unit vector $\vec{n}$ can be written as
\bea
n^\m = \f{k^\m}{u^\n k_\n} + u^\m \, ,
\eea
which satisfies $n^\m n_\m = 1$, $n_{(0)}=n_\m e^\m_{(0)}=n_\m u^\m = 0$ and $n_{(i)}=n_\m e^\m_{(i)}=k_{(i)}/k^{(0)}$ in the local rest frame of the fluid. Note that when calculating $n_{(i)}$, we have used the expression of the frequency, $\n=-k_\m e^\m_{(0)}=-k_\m u^\m=k^{(0)}$ in the local rest frame of the fluid. Then, it is convenient to define
\bea\label{nfApp}
f_{\m} \equiv  F_{\m\n}k^{\n}\,,
\eea
and evaluate it at the tetrad
\bea
f_{(0)} = F_{(0)(a)}k^{(a)} = - E_{(i)}k^{(i)} \, , \quad 
f_{(i)} = F_{(i)(a)}k^{(a)} = E_{(i)}k^{(0)} + \td{\varepsilon}_{ijk} B_{(k)} k^{(j)}\,,
\eea	
where $\td{\varepsilon}_{ijk}$ is the volume element of the three-dimensional spatial surface orthogonal to $u^\mu$. We can also obtain
\bea
f^{\m}f_{\m} = f_{(i)}f^{(i)}-f^2_{(0)} =\left[ \left(k^{(0)}\vec{E}+\vec{k}\times \vec{B}\right)^2-\left(\vec{k}\cdot\vec{E}\right)^2 \right] = \nu^2\mE^2 \,,
\eea
which is independent of the choice of reference frame. 

In addition, the absorption coefficient can be determined by Kirchhoff's law, 
\bea
\a_{\n} = \frac{j_{\n}}{B(T_\te,\n)}\,, \quad B(T_\te,\n) = \frac{2h\n^3}{\mathrm{e}^{h\n/k_BT_\te}-1}\,.
\eea
We consider the frequency of observed photons, denoted by $\nu_o$, to be less than $400$ GHz, while the temperature of the electron is estimated to be on the order of $10^{11}$ K. Henceforth, for the photons away from the horizon by more than $10^{-16}GM/c^2$, we find the ratio 
\bea
\frac{h\n}{k_BT_\te} = \frac{h\n_o}{g k_BT_\te} \sim \sqrt{-\left(1-\frac{2GM}{r c^2}\right)}\frac{h\n_o}{k_BT_\te} <0.02\,,
\eea
where $h$ is the Planck's constant and $g=\n_o/\n$ is the redshift factor. Here we employ the static emitter near a Schwarzschild black hole for simple estimation. This implies that the condition of $h\n \ll  k_B T_\te$ persistently holds. Thus, the radiative power adheres to the Rayleigh-Jeans law,
\bea\label{aeApp}
\a_{\n}\approx \f{j_{\n}}{2\nu^2 k_BT_\te} = \f{j_{\n}}{k_BT_\te}\f{g^2}{2\nu_o^2} \,.
\eea

\bibliographystyle{utphys}
\bibliography{fluidimage}
		
\end{document}